\documentclass[10pt, conference, letter]{IEEEtran}
\IEEEoverridecommandlockouts
\normalsize
\usepackage{cite}
\ifCLASSINFOpdf
\usepackage[pdftex]{graphicx}
\else
\fi
\usepackage{amsmath}
\usepackage{array}
\ifCLASSOPTIONcompsoc
\usepackage[caption=false,font=normalsize,labelfont=sf,textfont=sf]{subfig}
\else
\usepackage[caption=false,font=footnotesize]{subfig}
\fi
\usepackage{fixltx2e}
\usepackage{stfloats}
\usepackage{amssymb}
\usepackage{amsthm}
\usepackage{algpseudocode}
\usepackage{algorithm}

\usepackage{mathtools, cuted}
\usepackage{graphicx}
\usepackage{mathptmx}
\usepackage{amsmath}
\usepackage{xcolor}
\allowdisplaybreaks

\theoremstyle{definition}

\usepackage{setspace}
\makeatletter
\def\endthebibliography{%
	\def\@noitemerr{\@latex@warning{Empty `thebibliography' environment}}%
	\endlist
}
\makeatother

\begin{document}
\title{Dynamic Power Allocation and Virtual Cell Formation for Throughput-Optimal Vehicular Edge Networks in Highway Transportation 
}
	
\author{\IEEEauthorblockN{\textbf{Md Ferdous Pervej} and \textbf{Shih-Chun Lin}}
		\IEEEauthorblockA{Department of Electrical and Computer Engineering\\
			North Carolina State University, Raleigh, NC 27695, USA\\
			Email: \{mpervej, slin23\}@ncsu.edu 
			} 
\thanks{This work was supported by NC State 2019 FRPD, Cisco Systems, Inc., and North Carolina Department of Transportation (NCDOT).}}
\maketitle

\begin{abstract}
	This paper investigates highly mobile vehicular networks from users' perspectives in highway transportation. 
    Particularly, a centralized software-defined architecture is introduced in which centralized resources can be assigned, programmed, and controlled using the anchor nodes (ANs) of the edge servers. 
    Unlike the legacy networks, where a typical user is served from only one access point (AP), in the proposed system model, a vehicle user is served from multiple APs simultaneously.
    While this increases the reliability and the spectral efficiency of the assisted users, it also necessitates an accurate power allocation in all transmission time slots.
    As such, a joint user association and power allocation problem is formulated to achieve enhanced reliability and weighted user sum rate.    
    However, the formulated problem is a complex combinatorial problem, remarkably hard to solve. 
    Therefore, fine-grained machine learning algorithms are used to efficiently optimize joint user associations and power allocations of the APs in a highly mobile vehicular network.
	Furthermore, a distributed single-agent reinforcement learning algorithm, namely SARL-MARL, is proposed which obtains nearly identical genie-aided optimal solutions within a nominal number of training episodes than the baseline solution.
    Simulation results validate that our solution outperforms existing schemes and can attain genie-aided optimal performances.
\end{abstract}
	
	
	\IEEEpeerreviewmaketitle
	
	\section{Introduction}
	Vehicular networking has lately drawn significant attention in wireless networking, particularly for next-generation communication platform, such as beyond 5G systems. 
	The advent of vehicular ad-hoc networking \cite{4539481} has made it possible to establish communication among vehicles on the road. 
	This has posed as a direct method to share important safety messages among the vehicles using device-to-device like communication. 
	The early development of dedicated short-range communications (DSRC) \cite{jiang2006design} were designed following IEEE 802.11 physical and medium access layers technologies \cite{7497762}. 
	However, DSRC uses collision avoidance medium access schemes \cite{molina2017lte}. 
	Particularly, when network load increases, this creates an extra burden when extreme reliability has to be ensured \cite{araniti2013lte}.
	The so-called cellular vehicle-to-everything (C-V2X) was developed by the 3GPP as a potential alternative \cite{molina2017lte} to these limitations of DSRC.
	In C-V2X there are mainly two radio interfaces, namely the cellular interface (Uu interface) and the sidelink (PC5 interface). 
	VUs maintain their communication with the infrastructure using the Uu interface. 
	On the other hand, they can communicate with each other directly using the PC5 interface.
	The applications of V2X aim to incorporate both DSRC and C-V2X \cite{ghafoor2019enabling}, in reality, and have significantly drawn the attention recently to reduce travel time, traffic congestion, and ubiquitous internet connectivity for smart cities \cite{heo2019performance}. 
	
	There are several works in the literature addressing various aspects of vehicular networking. 
	{\c{S}}ahin \textit{et.al.} have presented a virtual cell formation for V2X in \cite{sahin2018virtual} where vehicles that are in close-proximity can receive downlink multicast data from the transmission points.
	Gao \textit{et. al.} have proposed a deep neural network based resource allocation scheme in \cite{li2019learning} where they have used block coordinated descent method and minimized weighted minimum mean square error.
	Lien \textit{et. al.} have considered a local area data network and shown a way to achieve lower latency in the radio access part in \cite{lien2018low}. 
	Considering the potential of performing caching at the edge nodes, the authors have used sophisticated tools of stochastic optimization and reinforcement learning (RL) to achieve lower latency using feedback less transmission. 
	He \textit{et. al.} have proposed an integrated platform for connected vehicles in \cite{8061008}. 
	The authors have formulated an integrated platform, in which they have orchestrated a joint networking, caching and computing resource allocation problem, and applied deep reinforcement learning (DRL) techniques. 	
	
	Tan \textit{et. al.} have incorporated both caching and computing in mobility aware vehicular networks in \cite{8447267}. 
	The authors have shaped vehicle mobility using a naive contact time-based model. 
	Considering both VUs and RSUs have caching and computing capabilities, they have used a DRL framework to find the optimal caching placement and computing resource allocation strategy in this paper. 
	Ye \textit{et. al.} have used DRL to allocate radio resources for vehicle-to-vehicle (V2V) communication in \cite{ye2019deep}. 
	In particular, they have reused the vehicle-to-infrastructure (V2I) uplink channels for V2V transmissions. 
	Distributed vehicular networking resource allocation problems have also been addressed in \cite{8792382} using multi-agent reinforcement learning. 
	Similar to Ye \textit{et. al.}, the authors have reused uplink frequency resources for V2V communications. 
	They have devised a system where their agents can interact with the environment and decide which channel to reuse and V2V transmission power level for V2V communication. 
	
\begin{figure}[t!]
		\centering
		\includegraphics[width = 0.5 \textwidth] {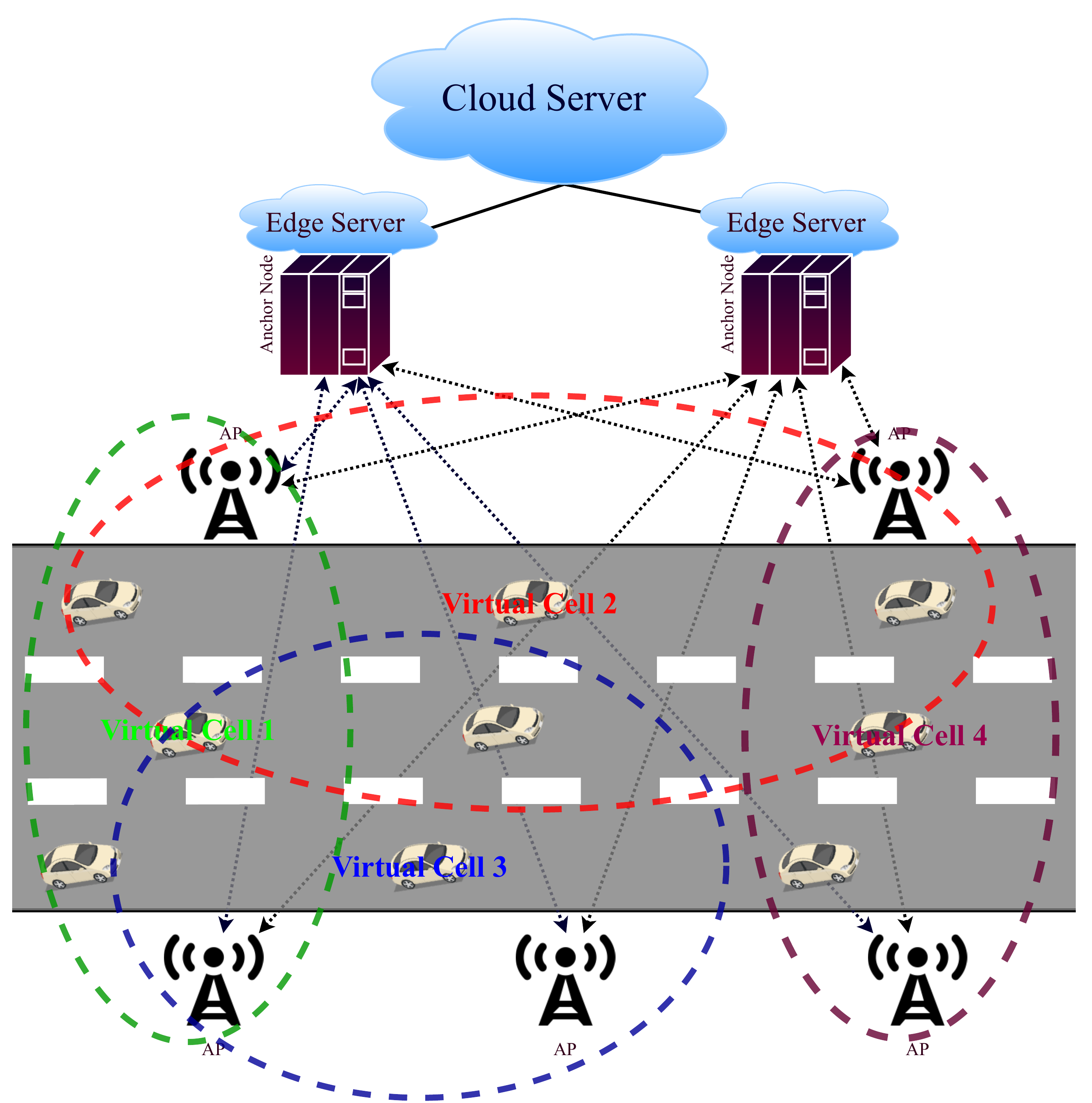} 
		\caption{User-centric vehicular edge networking.}
		\label{Sys_Mod}
	\end{figure}

	While most of these works address various aspects of V2X communication from the traditional network perspective, in this work, we incorporate user-centric dynamic cell design in a sophisticated software-defined environment to serve the users from multiple sources simultaneously.
	Particularly, we contemplate a centralized environment where resources can be programmed, controlled and distributed based on the users' demands.
	As presented in Fig. \ref{Sys_Mod}, we consider multiple edge servers are delicately deployed closer to the edge nodes in order to lessen end-to-end latency and improve spectral efficiency.
	Each of these edge servers is physically connected to a cloud server and controlled by its corresponding ANs.
	Furthermore, low-powered relay like APs are stationed as RSUs to guarantee ubiquitous connectivity of the VUs.
	Each of these APs is mesh connected to each of the edge servers.
	Moreover, we are interested in joint virtual cells design - for each of the scheduled VUs, and optimal beamforming vectors of the low-powered APs.
	Therefore, in this paper, we address joint user scheduling and power allocation problem in a highly mobile vehicular downlink network from the users' prospect.
	
	To the best of our knowledge, this is the first work to consider software-defined joint user-centric cell formation and power allocation, which ensures reliable communication and maximizing weighted sum rate (WSR) in a highway vehicular platform.
	Due to the complex combinatorial nature of the formulated optimization problem, it is challenging to solve the problem via traditional optimization methods.
	Thus, we propose a distributed Q-learning based RL solution to obtain optimal WSR and validate our results with genie-aided optimal, baseline single agent RL (SARL), multi-agent RL (MARL) and other baseline schemes.
	The rest of the paper is organized as follows. 
	Section \ref{system_model} presents our system model with vehicle node distributions, SDV2I communication and problem statements. 
	The proposed RL solution is described in Section \ref{RL_Approach} followed by results and discussions in Section \ref{Sim_Res}.
	Finally, Section \ref{Conclusion} concludes the paper.

	\section{System Model and Problem Formulation}
	\label{system_model}
	We present our software-defined system model, followed by the node distributions, communication model and dynamic user-centric cell formation problem in this section.
	
	\subsection{Software-Defined System Model}
	We consider highly mobile autonomous VUs are moving on the road. 
	Let us denote the VU set by $\mathcal{U} = \{u_1, u_2,\dots, u_U\}$, where $U \in \mathbb{Z}^+$. 
	Various low-powered APs are deployed in the surrounding geographic region. 
	Let the set of APs be denoted by $\mathcal{A} = \{a_1, a_2,\dots, a_A\}$, where $A \in \mathbb{Z}^+$. 
	The VUs move on the road and each VU establishes its connection with the network via these APs. 
	We assume that all of the APs are physically connected to edge servers.
	Note that, in reality, the number of such connections will be confined into a small group due to geographic locations of these edge nodes.
	Furthermore, each of the edge servers is controlled, programmed and operated by its respective AN.
	Let us denote AN by $b_l \in \mathcal{B}$. 
	In other words, there exists $|\mathcal{B}|$ numbers of edge servers each of which can be expressed by the identical notation of its AN.
	Each of the edge servers has a fixed and limited radio spectrum assigned by the cloud server.
	Let $W_l$ [Hz] denote the available radio resources of the edge server $b_l$.
	Considering an open-loop communication system, we assume that the channel state information (CSI) is perfectly known at each of the ANs.
	The ANs can form and schedule the beamforming weights of the APs for each of the virtual cells.
	Therefore, our system model is based on sophisticated SofAir~\cite{lin2018e2e}, where the ANs control, create and assign resource slices to the APs based on user's demands and thus, enhances the overall spectral efficiency.
	
	In order to ensure reliability and ubiquitous connectivity, we assume a VU can associate with multiple APs that are in its communication region.
	That is, in comparison to the conventional network-centric approach, we consider user-centric communication by formulating virtual cells for the scheduled users by associating each of them to multiple APs.
	Some patterns of forming virtual cells are shown by the dotted ellipses in Fig. \ref{Sys_Mod}. 
	Without any loss of generality, let us denote a user and an AP by $u_i \in \mathcal{U}$ and $a_j \in \mathcal{A}$, respectively.
	Due to a user-centric approach, let $\mathcal{A}_{i} (t)$ denote the set of APs that a VU $u_i$ can associate to, in time slot $t$.
	Furthermore, we denote the VU-AP association by the following indicator function:
	\begin{equation}
	\label{VU_AP_association}
	\begin{aligned}
	a_i^j(t) = \begin{cases}
	1, & \text{if AP $a_j$ is associated with VU $u_i$}\\
	0, & \text{otherwise} .
	\end{cases}
	\end{aligned}
	\end{equation}
	Moreover, an AP might be connected to multiple VUs at the same time.
	Let us further denote the set of VUs connected to an AP $a_j$ at time slot $t$ by the set $\mathcal{U}_{j}(t)$.
	We then have the following indicator function:
	\begin{equation}
	\label{AP_VU_association}
	\begin{aligned}
	u_j^i (t) = \begin{cases}
	1, & \text{if VU $u_i$ is associated with AP $a_j$}\\
	0, & \text{otherwise} .
	\end{cases}
	\end{aligned}
	\end{equation}
	Hence, $\mathcal{A}_{i}(t)$ denotes the set of APs that VU $u_i$ is connected to whereas, $\mathcal{U}_{j}(t)$ denotes the set of VUs connected to the AP $a_j$ in a given time slot.

	\subsection{Road Model and Node Distributions}
	\label{Road_model}
	We consider a straight three-lane one-way road structure without any intersection as our region of interest (ROI).
	Specifically, we consider the freeway case of 3GPP \cite{3gpp36_885}. 
	However, our modeling can be extended to complex practical road modeling.
	We are interested in establishing a communication framework for the V2X platform where the vehicle models are independent of road structure.
	The lane can be denoted by $L_m$, where $m \in \{1,2,3\}$ in our case. 
	We deploy VUs and APs following uniform distributions while maintaining a safety distance between two VUs.
	The method of updating mobility is described in Algorithm \ref{Road_Traffic_Model}. 
	Note that the Doppler effect has not considered in this paper and will be examined in our future work.
	
	\begin{algorithm} 
		\small
		\caption{Vehicular Road Traffic Modeling}
		\begin{algorithmic} [1]
			\State \textbf{Input}: Total number of VUs and lanes 
			\For {each time step, $t = 1,2,\dots,T_n$} 
			\For {each lane}
			\For {each vehicle}
			\State Update vehicle's position by adding linear displacement based on its velocity \Comment{ If new position is outside the ROI, TERMINATE}
			\EndFor
			\EndFor
			\State{\textbf{Return}: All vehicles' positions }
			\EndFor
		\end{algorithmic} \label{Road_Traffic_Model}
	\end{algorithm}

	\subsection{Software-Defined Vehicle-to-Infrastructure Communication}
	We assume that the VUs are equipped with a single antenna, whereas, each of the low-powered APs is equipped with a $N_j$ number of antennas.
	Unless mentioned otherwise, we assume an omnidirectional antenna for both VU and APs.
	Furthermore, with a slight abuse of notation, we represent the set of APs in $\mathcal{A}_{i}(t)$ by the set $\left\{a_{1,i}(t), a_{2,i}(t), \dots, a_{|\mathcal{A}_{j}|,i}(t) \right\}$, where $|\cdot|$ represents cardinality of set $\mathcal{A}_{i}(t)$.
	Moreover, $|\mathcal{A}_{i}(t)| \leq A$ i.e., the total number of APs assigned to the virtual cell $i$ of VU $u_i$ has to be less than or equal to the total number of available APs in the network for all time slot $t$.

	We model the wireless channel, between AP and VU, as an independent quasi-static flat fading model during a basic time block. 
	Let us denote the channel response at a VU $u_i$ from the AP $a_j$ by $\mathbf{h}_{a_j^i}(t) = [h_{ij_1}(t), h_{ij_2}(t), \dots, h_{ij_{N_j}}(t)]^T$, where $h_{ij_p}(t)$ denotes the channel between $u_i$ and the $p^{th}$ antenna of AP $a_j$ at time $t$.
	Then, the channel response at the VU $u_i$ from all APs is represented as follows: 
	\begin{equation}
	\label{ch_response}
	\begin{aligned}
	\mathbf{h}_{i}(t) &= \left[\mathbf{h}_{a_1^i}^T(t), \mathbf{h}_{a_2^i}^T(t), \dots, \mathbf{h}_{a^i_A}^T(t)  \right]^T 
	=  \mathbf{\mathrm{D}}_{i} (t) \rho_i(t) \tau_{i}(t) \in \mathbb{C}^{N \times 1},
	\end{aligned} 
	\end{equation}
	where $\mathbf{\mathrm{D}}_{i}(t)$, $\rho_i(t)$ and $\tau_{i}(t) \sim CN \left( \mathbf{0}, \mathbf{I}_{N}\right)$ are large scale fading, $log$-Normal shadowing and fast fading channel vectors, respectively. 
	Note that channels are modeled following appropriate measures listed in \cite{3gpp36_885}.
	Moreover, $N = \sum_{j=1}^A N_j$ denotes the total number of antennas in all APs.
	
	We assume linear downlink beamforming in our SD V2I platform. 
	Let us denote the beamforming vector for VU $u_i$ at AP $a_j$ by $\mathbf{w}_{a_j^i}(t) \in \mathbb{C}^{N_j \times 1}$ in time $t$.
	Then, the beamforming vector to VU $u_i$ can be denoted by $\mathbf{w}_i (t) \stackrel{\Delta}{=} \left[\mathbf{w}_{a_1^i}^{T}(t), \dots, \mathbf{w}_{a_A^i}^{T}(t)  \right]^T \in \mathbb{C}^{N_i \times 1}$.
	Furthermore, the entire network beamforming design can be denoted by $\mathbf{W}(t) \stackrel{\Delta}{=} \{\mathbf{w}_1(t), \dots, \mathbf{w}_U(t) \}$. 
	Now, let us also denote the unit powered intended signal for VU $u_i$ by $x_i(t) \in \mathbb{C}$.
	Therefore, we have to satisfy $\mathbb{E}[x_i^H(t) x_i(t)] = 1$ all the time.
	Furthermore, applying the beamforming vector, the transmitting signal of the AP $a_j$ is denoted as $\mathbf{s}_j(t) = \sum_{i = 1}^{U} \mathbf{w}_{a_j^i}(t) x_i(t)$. 
	Therefore, the received signal at the VU $u_i$ is then calculated as follows:
	\begin{subequations}
		\label{rx_signal_at_UE}
		\begin{align}
		&y_{i}(t) ~ = \sum_{a_j \in \mathcal{A}} \mathbf{h}_{a_j^i} ^H (t) \mathbf{s}_j(t)  + \eta_i(t) \\
		&= \underbrace{\mathbf{h}_{i}^H(t) \mathbf{w}_i(t) x_i(t)}_\text{desired signal} + \underbrace{\sum_{u_{i'} \in \mathcal{U}\backslash u_i} \mathbf{h}_{i}^H(t) \mathbf{w}_{i'}(t) x_{i'}(t)}_\text{interference}  + \underbrace{\eta_i(t)}_\text{noise}, 
		\end{align}
	\end{subequations}
	where $\eta_i(t) \sim CN\left(0, 1 \right)$ denotes the received noise which is zero mean circularly symmetric Gaussian distributed with variance $\sigma^2$. 
	
	\subsection{User-Centric Dynamic Cell Formation}
	\label{UE_Centric_Cell}
	
	With our analysis and vehicular traffic modeling, as presented in \ref{Road_model}, we now aim to derive the instantaneous achievable rate at the VU. 
	At first, we calculate the signal-to-interference-plus-noise ratio (SINR) as follows:
	\begin{equation}
	\label{SINR}
	\gamma_{i}^t\left(\mathbf{W}(t)\right) = \frac{\left| \mathbf{h}_{i}^H(t) \mathbf{w}_{i}(t) \right|^2}{\sigma^2 + \sum_{u_{i'} \in \mathcal{U} \backslash u_i}  \left|\mathbf{h}_{i}^H(t) \mathbf{w}_{i'}(t) \right|^2}
	\end{equation}
	Considering time division duplexing (TDD) operated system, we thus calculate the instantaneous achievable data rate for VU $u_i$ as follows:
	\begin{equation}
	\label{ergodic_data_rate}
	R_i^t \left(\mathbf{W(t)}\right) = \left(1-\kappa \right)\log_2 \left(1 + \gamma_{i}^t \left(\mathbf{W}(t)\right) \right),
	\end{equation}
	where $\kappa$ represents spectral efficiency loss due to signaling at the APs.
	Note that if a user is scheduled during the transmission time slot, the beamforming vector $\mathbf{w}_i(t)$ is nonzero, thus the rate is nonzero as well.
	On the other hand, if a user is not scheduled, the beamforming vector for that user should be zero leading to a zero achievable rate.
	Furthermore, as multiple APs are serving each of the scheduled users, the backhaul resource consumption by each of those users should also be carefully calculated.
	As such, next, we calculate the backhaul resource consumption\cite{6831362} as follows: 
	\begin{equation}
	\label{backhaul_consump_u_i}
	\mathrm{C}_i(t) = \left \Vert \left[ \left \Vert \mathbf{w}_{a_1^i}(t) \right\Vert_2, \dots, \left \Vert \mathbf{w}_{a_A^i}(t) \right\Vert_2\right]  \right \Vert_0 R_i^t \left(\mathbf{W}(t)\right),
	\end{equation}
	where $\left \Vert \cdot \right \Vert_0$ represents the total number of nonzero elements in a vector and commonly referred as $l_0$-norm.
	
	Now, due to resource constraints, we may not serve all users at the same time.
	Therefore, there will be certain restrictions on the number of active users in the VU set $\mathcal{U}$.
	However, recall that our formulated rate calculation in equation (\ref{ergodic_data_rate}) can also be used to design the user scheduling.
	We intend to serve all active users in $\mathcal{U}$ in a transmission time slot by forming virtual cells for each of the users and dynamically selecting the transmission power of the APs.
	We aim to find optimal user-centric cell formation and beamforming weights calculation for the APs in our objective function.
	The question that we try to answer is - \textit{what are the optimal user associations and APs transmission powers that maximize the throughput in our SD controlled highly mobile vehicular network?}
	A naive approach would be serving a user from as many APs as possible with the maximum transmission powers of the APs.
	However, transmitting to a particular VU from the APs with maximum transmission power will severely impact the SINR performances of the other active VUs.
	Therefore, it is essential to know the optimal transmission power of the APs for each of the active VUs.
	
	We present our joint optimization problem as follows:
	\begin{subequations}
		\label{Original_Problem}
		\begin{align}
		& \textbf{Find:}  && a_i^j (t),  u_j^i(t),  \mathbf{w}_{a_j^i}(t), ~~ \forall i \in \mathcal{U}, j \in \mathcal{A}  \nonumber \\ 
		\label{OP1}& \textbf{Maximize} && \sum_{i \in \mathcal{U}}  \zeta_{i}(t) \mathrm{C}_i(t) \\
		\label{Cons_1} & \textbf{Subject to} && 1 \leq |\mathcal{A}_{i}(t)| \leq A,~~ \forall i \in \mathcal{U}\\
		\label{Cons_3} &~&& \gamma_{i}^t\left(\mathbf{W}(t)\right) \geq \gamma_i^{min}, ~~\forall i \in \mathcal{U}\\
		\label{Cons_4} &~&& \sum_{i = 1}^U \left \Vert \mathbf{w}_{a_j^i}(t) \right \Vert_2^2 \leq P_j^{max},~~ \forall j \in \mathcal{A}\\
		\label{Cons_9} &~&& a_i^j(t) \in \{0,1\}, u_j^i(t) \in \{0, 1\},
		\end{align}
	\end{subequations}
	where $\zeta_{i}(t)$, $\gamma_i^{min}$ and $P_j^{max}$ are the weights of the data rate of user $u_i$ at time $t$, minimum SINR requirement for reliable communication and maximum allowable transmit power of AP $a_j$, respectively. 
	Note that constraint (\ref{Cons_1}) ensures that our user-centric virtual cell contains more than one AP to form the cluster. 
	Constraint (\ref{Cons_3}) ensures the obtained SINR is greater than a minimum threshold.
	Moreover, constraint (\ref{Cons_4}) limits the total transmit power of AP $a_j$ to be at maximum $P_j^{max}$. 
	
	Due to the $l_0$ norm, the formulated problem is not suitable to be solved using a gradient-based algorithm.
	Furthermore, due to the combinatorial nature of the originated problem, it is extremely hard to solve within a short period.
	Note that if we know all $a_j^i(t)$s - $\forall j \in \mathcal{A}$, we can figure out the $u_j^i(t)$s using equations (\ref{VU_AP_association}-\ref{AP_VU_association}).
	Therefore, for each of the AP, there are $2^U-1$ possible combinations, only for the VU-AP associations, in a single time-slot. 
	Besides, for each of these associations, the AP has to select the power level for each of the active users. 
	Moreover, in a centralized controlled environment, the centralized agent needs to make a central decision for all such AP-VU associations and power levels selection pairs.
	As such, it is obvious that as the number of APs and VUs increases the complexities increases exponentially.
	Therefore, we employ an elegant machine learning approach to solve the problem in what follows.
	
	\section{Reinforcement Learning-Based Vehicular Edge Slicing Mechanism}
	\label{RL_Approach}
	In this section, we discuss the problem-solution from the RL perspective.
	We first clearly state the state, action, and reward of the RL agent.
	Then we present our learning algorithms.
	
	\subsection{State}
	The state-space contains all channel state information $\mathbf{H}^t$.
	It also contains the geographic locations of the VUs and APs.
	Let us denote the positions of the VUs and APs at state $\textbf{s}_t$ by $\mathbf{X}_i^t$ and $\mathbf{X}_j^t$, respectively.
	Therefore, we denote our state space as follows:
	\begin{equation}
	\label{state}
	\mathbf{s}_t = \left\{ \mathbf{X}_i^t, \mathbf{X}_j^t, \mathbf{H}^t \right\}.
	\end{equation}
	
	\subsection{Action}
	The action space contains the VU-AP association indicator functions followed by the beamforming vectors for the selected VU-AP associations.
	\begin{equation}
	\label{action}
	\mathrm{\mathbf{a}}_t = \left\{ a_j^i(t) ~\forall i \in \mathcal{U} , \mathbf{w}_{a_j^i}(t) ~\forall j \in \mathcal{A} \right\} 
	\end{equation}
	Notice that, in equation (\ref{action}), the RL agent needs to take two action sets i.e., VU-AP associations and beamforming weights. 
	However, both of these actions cannot be performed at the same time.
	Simply put, for deciding the optimal beamforming weights, the agent needs to know the VU-AP associations.
	Therefore, the action set can be thought of as a twin step process.
	At first, the RL agent needs to decide the VU-AP associations.
	Based on that decision, it then can allocate each AP's transmission power for all of the VU that are being served by the respective AP.
	Therefore, we divide the action space into twin scales as described below.
	
	Considering an open-loop system - one-shot transmission, to ensure reliability and increase the user data rate, we have considered serving a scheduled VU from multiple APs.
	If a user is scheduled to be served, we then design the beamforming vectors of the APs for that particular user.
	Now, note that we assume perfect CSI is known at the AN.
	As such, we model the beamforming weights using the following equation:
	\begin{equation}
	\label{beam_power}
	\mathbf{w}_{a_j^i}(t) = \frac{\mathbf{h}_{a_j^i}(t)}{\left \Vert \mathbf{h}_{a_j^i}(t) \right \Vert_2} \times \sqrt{P_{a_j^i}(t)},
	\end{equation}
	where $\mathbf{h}_{a_j^i}(t)$ is the wireless channel information from AP $a_j$ to VU $u_i$ and $P_{a_j^i}(t)$ is the allocated transmission power of AP $a_j$ to transmit to VU $u_i$.
	Therefore, we rewrite the action space as: 
	\begin{equation} 
	\mathrm{\mathbf{a}}_t = \left\{ a_j^i(t) ~\forall i \in \mathcal{U} , P_{a_j^i}(t) ~\forall i \in \mathcal{U}~\& ~j \in \mathcal{A} \right\}.
	\end{equation}
	Now, instead of continuous power level, we divide the APs transmission power level into multiple discrete levels.
	Particularly, we divide $P_{a_j^i}(t)$ into $4$ levels (\textit{e.g.}, $P_{a_j^i}(t) \in \{5, 10,15,20\}$ dBm). 
	In other words, each AP can select its transmission power to serve a user from one of these power levels. 
	Therefore, our objective function is still the same as presented in equation (\ref{Original_Problem}).
	We formulate the beamforming vectors using the optimal power selection and equation (\ref{beam_power}).

	\subsection{Reward}
	Without any loss of generality, we define the weighted sum rate as the reward function for our learning algorithm. We also ensure that each of the users receives the minimum SINR threshold for a chosen action; otherwise, we return a zero reward. 
	Accordingly, the reward function is given as follows:
	\begin{equation}
		r_t = \begin{cases}
		\sum_{i \in U} \zeta_{i}(t)\mathrm{C}_i(t), & \text{if } \gamma_{i}^t\left(\mathbf{W}(t) \right) > \gamma_i^{min} ~\forall i \in \mathcal{U}\\
		0, & \text{otherwise}.
		\end{cases} 
	\end{equation}
	\subsection{Single Agent Reinforcement Learning (SARL)}
	$Q$-learning is a model-free RL framework \cite{watkins1992q} which takes the state and action into account and solve hard optimization problem such as equation (\ref{Original_Problem}) efficiently.
	In each state $\mathbf{s}_t$ the agent takes an action $\mathbf{a}_t$ from which it gets a reward $r_{t}$ and the environment transit to the next state $\mathbf{s}_{t+1}$. 
	The governing equation of Q-learning is shown in the following: 
	\begin{equation}
	\label{q_table}
	Q(s_t,a_t) \!\leftarrow \!\! (1-\alpha) Q(s_t, a_t) + \alpha \left(r_t + \gamma \underset{\textbf{a}}{\text{ max }} Q(s_{t+1},\textbf{a})\right)\!,
	\end{equation} 
	where $\alpha$ and $\gamma$ are learning rate and discount factor, respectively. 
	Our learning algorithm for SARL is presented in Algorithm \ref{Q_learning}.
	
	\begin{algorithm} 
		\small
		\caption{$Q$ - learning based RL Algorithm}
		\begin{algorithmic} [1]
			\State \textbf{Initialize}: For all possible states and actions generate random $\mathbf{Q}(\mathbf{s}_t, \mathbf{a}_t)$ table
			\For {each episode}
			\State Initiate the environment 
			\State Generate $	\mathbf{s}_t = \left\{ \mathbf{X}_i^t, \mathbf{X}_j^t, \mathbf{H}^t \right\}$
			\While {not terminated}
			\State Take action using $\epsilon$-greedy policy
			\State Calculate reward, $r_t$, transit to next state $\mathbf{s}_{t+1}$
			\State Update $Q$ table following equation (\ref{q_table})
			\State $\mathbf{s}_t \leftarrow \mathbf{s}_{t+1}$
			\EndWhile \Comment {If $\mathbf{s}_t$ is the terminal state}
			\EndFor 
		\end{algorithmic} \label{Q_learning}
	\end{algorithm} 
	
	\subsection{Multi-agent Reinforcement Learning (MARL)}
	In MARL, multiple agents can take their individual actions and get an overall centralized reward for their joint decisions. 
	Liu \textit{et. al.} have recently presented a MARL framework in \cite{liu2019trajectory} in which each agent has its own $Q$-table and can take action independently.
	Particularly, following the $\epsilon$-greedy policy, if the value of the random variable is greater than $\epsilon$, the authors have chosen the action of the $n$th agent using the following equation: 
	\begin{equation}
	a_n^t = \text{arg max}_a \left(\sum_{1 \leq j \leq N} Q_j^t(s_j^t, a) \right),
	\end{equation} 
	where $j = 1,\dots, N$ represents the cooperating agents.
	The update rule for the $n$th agents follows the following equation \cite{liu2019trajectory}: 
	\begin{equation}
	\!\!Q_n(s_n,a_n) \!\! \leftarrow \!\! (1-\alpha)Q_n(s_n,a_n) + \alpha \left(\!\! r_n(s_n, \bar{a}) + \gamma \text{ max} \underset{ b \in A_n}{Q_n}(s'_n, b) \!\!\! \right).
	\end{equation} 
	
	\subsection{Distributed SARL with Multi-agent Learning (SARL-MARL Collaboration)}
	While SARL is the baseline RL scheme, it may suffer to attain the best performance if the action and state space are too large.
	In such a platform MARL can be used to shrink the number of action space.
	However, whether MARL will achieve the optimal performance is still questionable as each agent takes its decision independently. 
	Although, Liu \textit{et. al.} \cite{liu2019trajectory} have considered collaboration among the agents, we validate through simulation results that this scheme fails to attain the best performance in our specific problem.
	As an alternative, we incorporate the concept of distributed learning. 
	Since a single agent RL framework has to evaluate all possible actions, we consider dividing the action space into multiple agents.
	We use individual $Q$-table and keep track of the global best performance.
	Note that the number of possible states for all agents are still the same as SARL, we are only evaluating the performance from segmented action spaces.
	
	Let us denote the number of agents by $N$. 
	Then, the dimension of the $Q$-table for each of the agents is $\mathbb{R}^{S \times A/N}$, where $S$ and $A$ represents the size of the state space and action space, respectively.
	Note that we selected $N$ such a way that $A/N \in \mathbb{Z}^+$.
	In each of the state $\mathbf{s}_t$, each of these $N$ agents takes their action following the same methodology as the baseline $Q$-learning.
	Let us denote $\textbf{Q}_\text{central} \in \mathbb{R}^{S}$ a centralized vector that stores the global best action in each of the states.
	In each of the time step, we update the $\textbf{Q}_\text{central} $ using the following equation: 
	\begin{equation} 
	\label{Q_central_update}
	\textbf{Q}_\text{central} [\textbf{a}_t] \leftarrow \begin{cases}
	\textbf{Q}_\text{central} [\textbf{a}_t], ~\text{if any } r_t [\textbf{a}_t] > r_t [\textbf{a}_t^{\text{old}}], \forall \text{agents} \in N\\
	\textbf{Q}_\text{central} [\textbf{a}_t^{\text{old}}], \text{ otherwise}.
	\end{cases}
	\end{equation}
	Note that each of the agents follows and updates its action and $Q$-table according to the baseline SARL.
	The detailed procedure of our distributed SARL with multi-agent learning is presented in Algorithm \ref{SARL_MARL}.
	
	\begin{algorithm} 
		\small
		\caption{SARL-MARL: Distributed SARL}
		\begin{algorithmic} [1]
			\State \textbf{Initialize}: Total number of agents, $N$, divide the large action space into $N$ small sets. 
			\State For all possible states and action sets generate random $\mathbf{Q}_l(\mathbf{s}_t, \mathbf{a}_t)$ tables, where $l \in N$ represents the agent id.
			\State Generate $\textbf{Q}_\text{central} \in \mathbb{R}^S$ randomly
			\For {each episode}
			\State Initiate the environment, generate $	\mathbf{s}_t = \left\{ \mathbf{X}_i^t, \mathbf{X}_j^t, \mathbf{H}^t \right\}$
			\While {not terminated}
			\For {each $l \in N$}
			\State Observe the environment; choose $\textbf{a}_t$, based on the observation, following $\epsilon$-greedy policy; receive reward $r_t$; update its $Q$-table using equation (\ref{q_table})
			\If {$r_t >$ reward using $\textbf{Q}_\text{central}[\textbf{a}_t]$ }
			\State update $\textbf{Q}_\text{central}$ using equation (\ref{Q_central_update})
			\EndIf
			\EndFor
			\State $\mathbf{s}_t \leftarrow \mathbf{s}_{t+1}$
			\EndWhile \Comment {If $\mathbf{s}_t$ is the terminal state}
			\EndFor 
		\end{algorithmic} \label{SARL_MARL}
	\end{algorithm}  
	
	\section{Performance Evaluation}
	\label{Sim_Res}
	We consider $500$ meters of a three-lane one-way freeway as our ROI. 
	Vehicle's new position is generated by adding linear displacement with VU velocity of $140$ km/h as per \cite{3gpp36_885}. 
	The channel model and related parameters are also chosen as specified in \cite{3gpp36_885}. 
	In order to keep a tractable sate space, we consider updating it after every $100$ milliseconds.
	Furthermore, for the ease of simulation, we consider a full buffer network model in which each of the APs serves all VUs simultaneously.
	Note that our proposed problem solution can work in other scheduling algorithms as well. 
	As we are not adopting any actual scheduling schemes, we consider full buffer scenarios with $\zeta_{i} (t) = 1,~\forall i \in \mathcal{U} \& ~t$.
	We consider $1$ AN, $3$ APs and $3$ users. 
	While the VUs are dropped uniformly in each lane, the APs are placed $150$ meters apart fixed locations.
	We have assumed $8$ antennas per AP.
	For a tractable state space, we have considered that, at a given time step, all VUs are in the same $x$ locations - while they have different $y$ locations.
	We also have $\gamma_i^{min} = 10$ dB $\forall i\in \mathcal{U} \&~ t$ and $\kappa = 0.1$. 
	
	\begin{figure*}[t] 
		\centering
		\subfloat[Performance comparisons of different schemes] {\label{Rewards_Comparision} \includegraphics[width = 0.5 \textwidth]{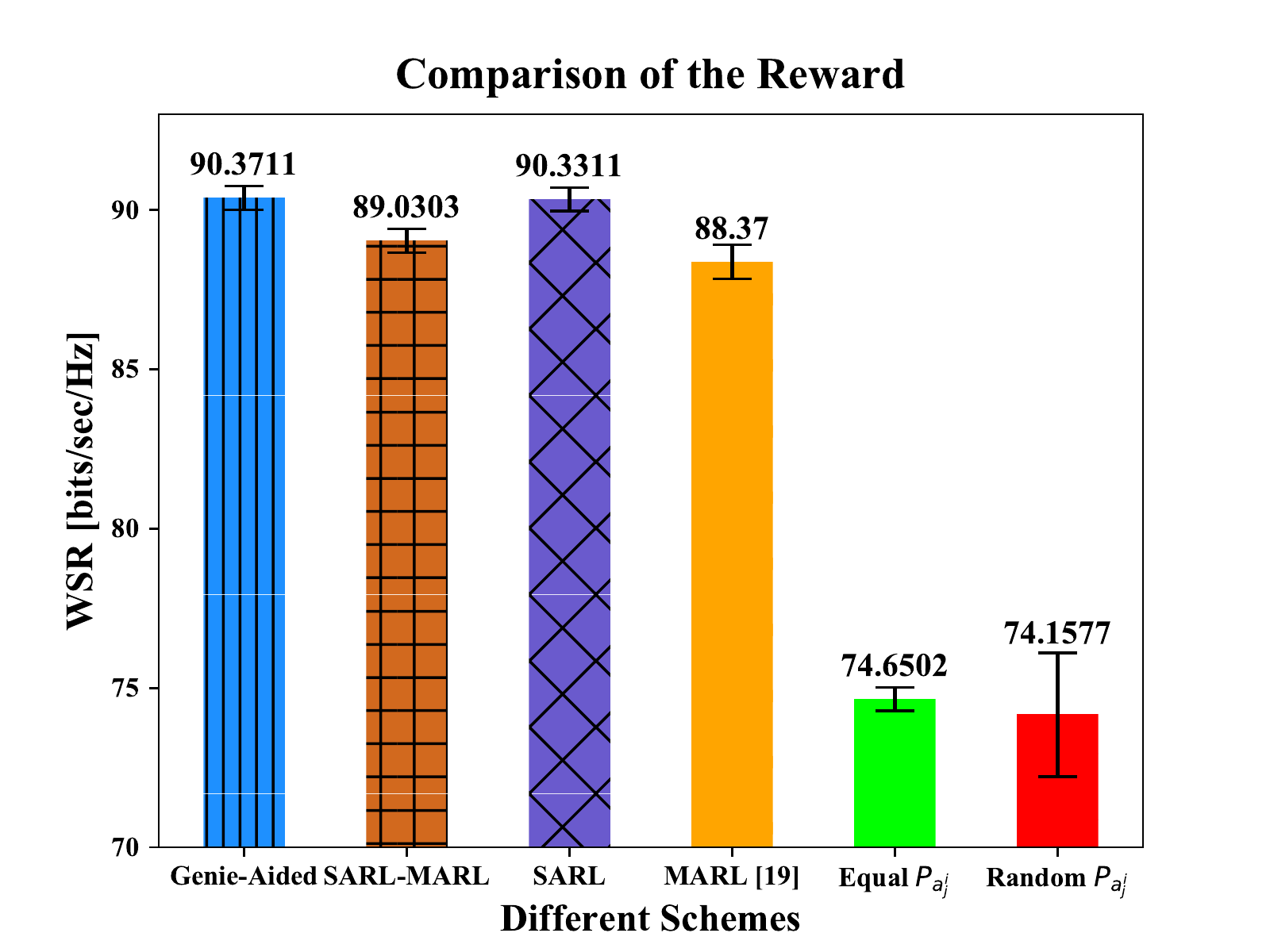} } 
		\subfloat[User fairness] {\label{UE_Fariness} \includegraphics[width = 0.5 \textwidth]{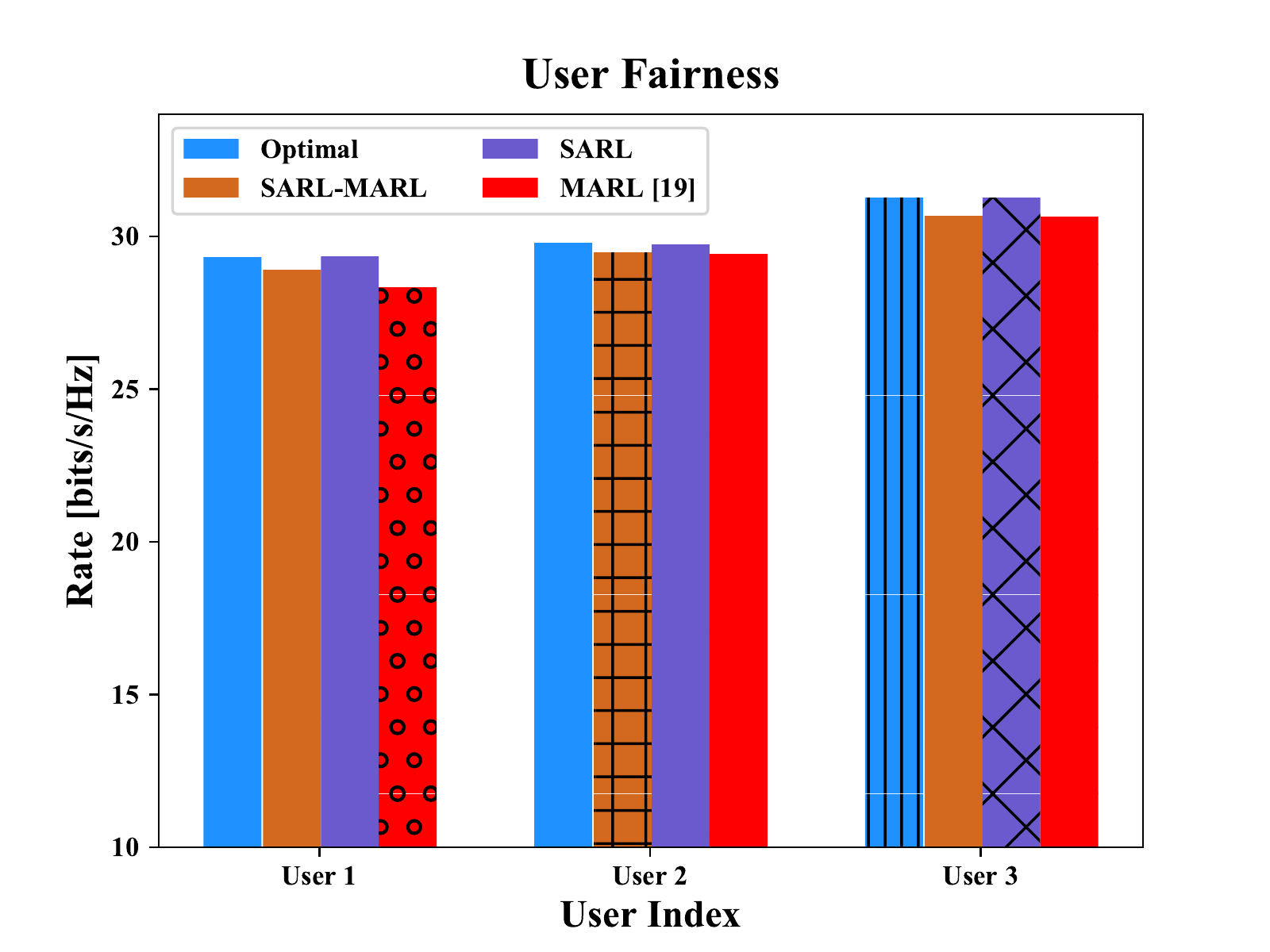}}
		\caption{Performance comparisons: AP coverage radius = $250$ m, $\gamma_{i}^{min}=10$ dB.} 
		\label{Performance_Comparision}
	\end{figure*}
	
	For the SARL, MARL and SARL-MARL collaborative algorithms, we have considered $\gamma = 0.8$.  
	Besides, the value of both $\epsilon$ and $\alpha$ are decayed linearly from $1$ to $0.01$ in each episode.
	Also, we have considered $3$ APs as independent agents for the MARL algorithm of Liu \textit{et. al.} \cite{liu2019trajectory} and we have used $N=4$ for our SARL-MARL distributed learning algorithm.
	For both SARL and MARL cases, we have trained our models for $1 \times 10^5$ episodes while we only train our model in $2.5 \times 10^4$ episodes in the SARL-MARL distributed learning case.
	In order to validate the results of each of these algorithms, we have compared the results with the \textit{genie-aided: optimal} solution.
	We have also used two baseline schemes, namely, (1) \textit{random power}: each AP randomly choose its transmitting power from $\{5,5,15\}$ and (2) \textit{equal power}: each AP transmit to all VUs using $P_j^{max}/U$ dBm power.
	
	We compare the average rewards of all of these schemes after running $250$ test episodes in Fig. \ref{Rewards_Comparision}.
	Our proposed SARL-MARL, baseline SARL, and MARL algorithms deliver average per VU reward of $29.68$, $30.11$, and $29.46$ bits/sec/Hz.
    These are nearly identical to the genie-aided optimal per VU reward of $30.12$ bits/sec/Hz.
    However, recall that we have trained both SARL and MARL for $10^5$ episodes, whereas the proposed SARL-MARL distributed learning algorithm has been trained on only $2.5 \times 10^4$ episodes.
	The performance of the equal and random power allocation schemes are, as expected, worse than the RL schemes.
	With regard to user fairness, our RL schemes and reward function always ensure that each of the users receives a fair data rate.
    Since we return zero rewards in the case that any of the user's SINR is below the threshold level, we expect that our RL agents learn this pattern quickly.
	This is also verified by our simulation results in Fig. \ref{UE_Fariness}.
	
\begin{figure*}
		\centering
		\subfloat[SINR threshold vs. WSR] {\label{SINR_Min_vs_WSR} \includegraphics[width = 0.5 \textwidth]{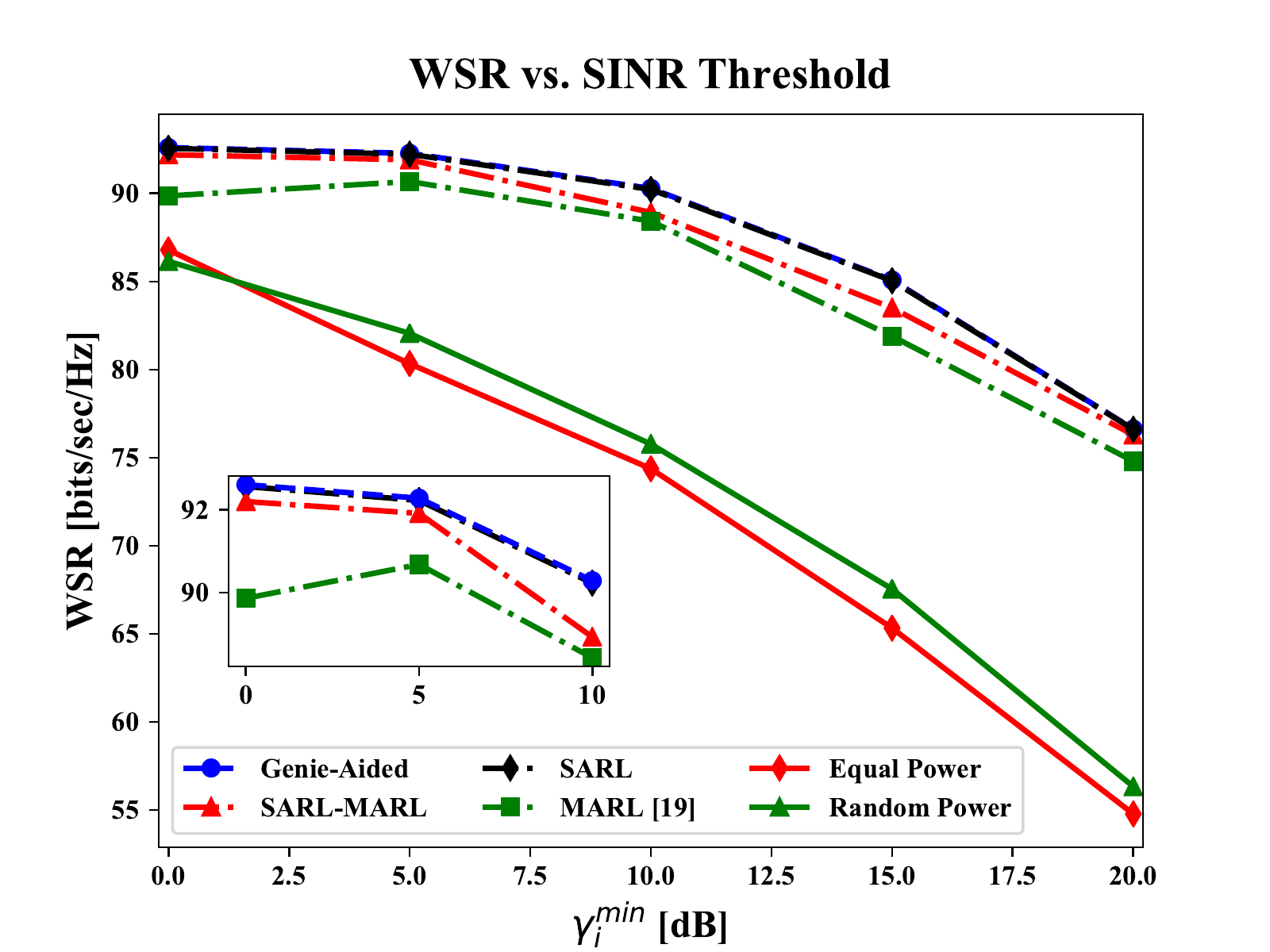} 
		}
		\subfloat[SINR threshold vs. success probability] {\label{Succ_Pro} \includegraphics[width = 0.5 \textwidth]{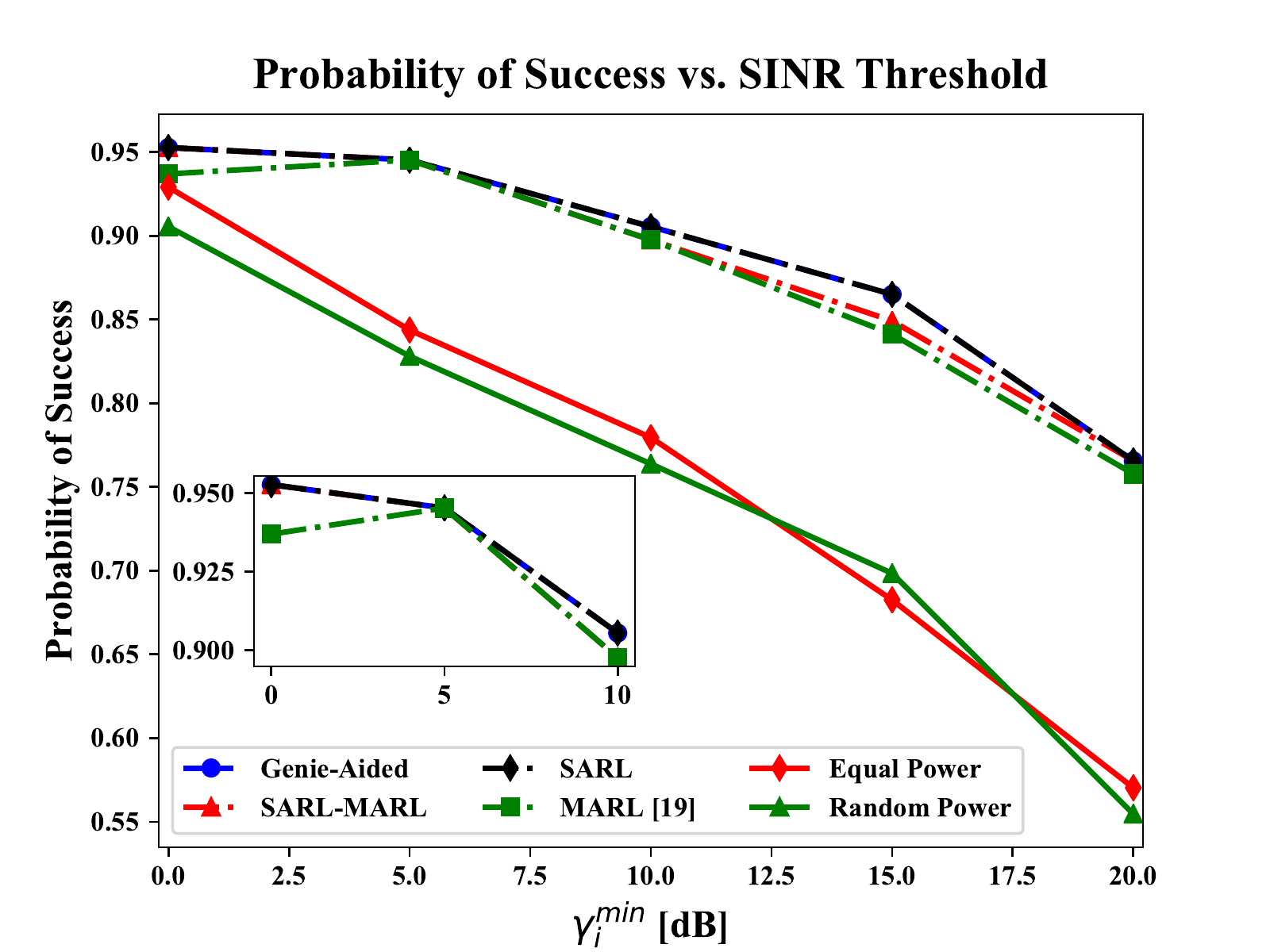}}
		\caption{Impact of the SINR threshold.} 
		\label{sinr_impact}
	\end{figure*}

	We further consider the impact of the reliability threshold.
	Note that, in our optimization problem, we have set the reliability constraint that each of the users has to get a minimum SINR threshold to ensure reliability.
	This, therefore, guarantees that regardless of the environment, our RL will learn to allocate resources such a way that each of the VUs receives at least $\gamma_{i}^{min}$ SINR.
	Furthermore, Fig. \ref{sinr_impact} indicates that as this threshold increases, the network should experience difficulties in achieving this demand for all of the users.
	Both Figs. \ref{SINR_Min_vs_WSR}-\ref{Succ_Pro} confirm that our proposed distributed SARL multi-agent learning (SARL-MARL) algorithm reaches near-optimal results compare to that of Liu \textit{et. al.}'s MARL and other baseline schemes.
	The performance gap between the optimal and the proposed SARL-MARL results are very close.
	Besides, the gap between our proposed solution and other baseline schemes are hugely evident as the reliability threshold increases.
	
\begin{figure}
		\centering
		\includegraphics[width = 0.5 \textwidth]{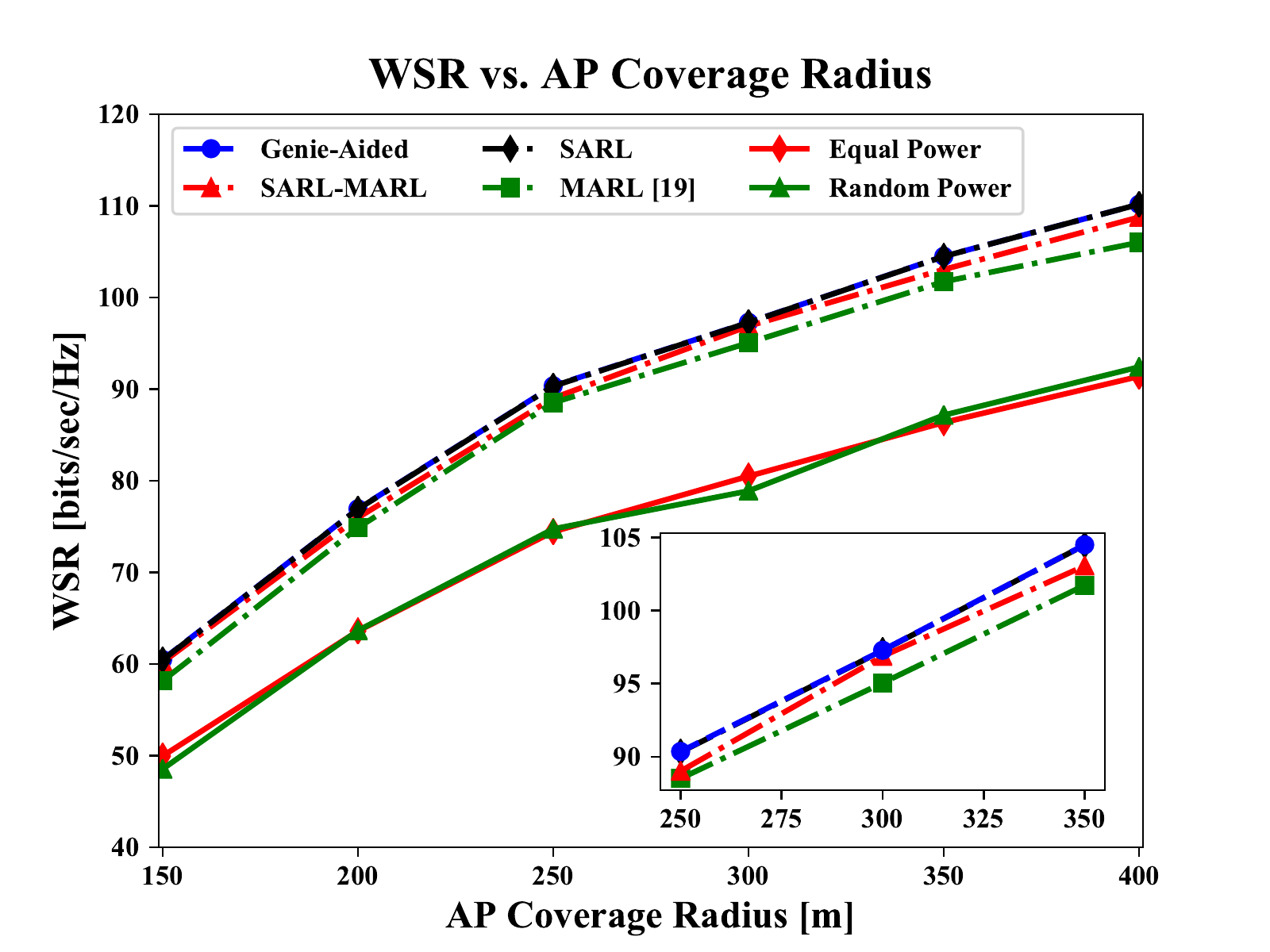} 
		\caption{Impact of the coverage radius of the access point.} 
		\label{Rewards_Vs_Cov_Radius}
	\end{figure}

	Finally, we show the impact of the coverage radius of the APs in Fig. \ref{Rewards_Vs_Cov_Radius}.
	Note that a VU can only be served if its in the coverage region of the AP.
	Therefore, as the coverage radius of the AP increases, more VUs can be served by that AP.
	As such, the expected sum rate of the user should increase, if the power levels of the APs are chosen appropriately with the increase of its coverage radius.
	This is also reflected and validated in our result presented in Fig. \ref{Rewards_Vs_Cov_Radius}. 
	
	\section{Conclusion} 
	\label{Conclusion}
	In this paper, we have presented an efficient way to dynamically allocate the transmission power of the APs and virtual cell formation for the VUs in user-centric vehicular networks.
	Using well-bred machine learning algorithms, we have demonstrated that the original hard combinatorial problem can be solved efficiently.
	Furthermore, as the numbers of possible states and actions increase, the traditional SARL suffers to behave optimally due to the curse of dimensionality.
	While MARL requires quite a large number of training episodes to attain near-optimal performance, the proposed SARL-MARL can achieve similar performance within a nominal number of training episodes. 
	
	\bibliography{Reference.bib}
	\bibliographystyle{IEEEtran}
	
\end{document}